\documentclass[preprint,12pt]{elsarticle}

\usepackage{amssymb}
\usepackage{amsmath}
\usepackage{parskip}
\usepackage{gensymb}
\usepackage{lipsum}
\usepackage{caption}
\usepackage{multirow}

\usepackage{pdfpages}
\usepackage{graphicx}
\usepackage{float}

\journal{arXiv}

\begin{document}

\begin{frontmatter}



\title{Adaptation of the visibility graph algorithm to find the time lag between hydrogeological time series}


\author{Rahul John$^{a, *}$, Majnu John$^{b, c, *}$\footnote{$* = $ contributed equally \\ Corresponding author: Department of Mathematics, $308$ Roosevelt Hall, $130$ Hofstra University, Hempstead, NY 11549.  e-mail: {\sf Majnu.John@hofstra.edu}}}

\address{$^{a}$Water Science Associates, Inc., \\Fort Myers, FL.}
\address{$^{b}$Department of Mathematics, \\ Hofstra University, \\Hempstead, NY.}
\address{$^{c}$The Feinstein Institute of Medical Research, \\Northwell Health System, \\Manhasset, NY.}

\begin{abstract}

Estimating the time lag between two hydrogeologic time series (e.g. precipitation and water levels in an aquifer) is of significance for a hydrogeologist-modeler. In this paper, we present a method to quantify such lags by adapting the visibility graph algorithm, which converts time series into a mathematical graph. We present simulation results to assess the performance of the method. We also illustrate the utility of our approach using a real world hydrogeologic dataset.

\end{abstract}

\begin{keyword}


time series, visibility graph algorithm, hydrogeology, aquifer water level, precipitation

\end{keyword}

\end{frontmatter}





\noindent
\section{Introduction}

In the field of Hydrogeology, many interesting concepts are related to finding the lag between two time series. For example, it is often hypothesized that for a seepage lake there is a significant time lag between net precipitation (precipitation minus water loss through evaporation and runoff) and the water levels over time, while such a lag for a drainage lake is often nonexistent or insignificant.  Seepage lakes are hydraulically isolated from surface water features and primarily fed by groundwater and direct precipitation.  Drainage lakes are typically connected to a network of streams and rivers (Wisconsin Department of Natural Resources, 2009).

Another example, which is our motivating example, is the relationship between precipitation and water levels of a shallow well in an unconfined aquifer versus water levels in a relatively deeper well in a semi-confined aquifer.  This relationship is particularly important to water resource managers and groundwater modelers who need to accurately quantify groundwater recharge into aquifers, for developing water-supply-plans for sustainable use of aquifers.  Groundwater recharge, defined as entry of water into the saturated zone, is influenced by a wide variety of factors including vegetation, topography, geology, climate, and soils (Dripps, 2003, Dripps, Hunt and Anderson 2006). Groundwater recharge, which is a small percentage of the precipitation that eventually reaches the water table, is one of the most difficult parameters to quantify.  This is because processes such as evaporation, transpiration and infiltration through unsaturated subsurface must first be estimated to determine the amount of water lost after a rainfall event.  Often times, groundwater models are developed by estimating the groundwater recharge using empirical relationships or as a percentage of precipitation.  It is a common practice to use groundwater recharge as a calibration parameter, meaning the recharge value that provides the best calibration to the model is selected as representative for the watershed simulated.  For temporal simulations, the lag time between a rainfall event and groundwater recharge into deeper aquifers are often ignored.

Although the underlying hydrogeological theory supports the existence of above time lags between time series, evidence based on empirical data for such lags have been typically assessed using visual inspection (e.g. Westoff \textit{et al}, 2010 in a different hydrogeological context) or cross-correlations (Levanon \textit{et al}, 2016) in hydrogeological literature. Cross-correlation method is essentially a parametric method, where certain parameters has to be estimated under the transfer-function-model framework and certain assumptions (such as joint bivariate stationarity of the two time series) has to be met (see Chapter 14, Wei 2006). Also diagnostic checking for model adequacy (such as whether the noise series and the input series are independent - see again Chapter 14, Wei 2006 for the definition of the noise series and input series referred to) has to be done before cross-correlograms are plotted, although such checking are rarely done in practice. In this paper, we propose a non-parametric method to quantify the time lag using a simple adaptation of the visibility graph algorithm (VGA), which is an algorithm that converts a time series into a graph and was developed by physicists and seen mainly only within the physics literature so far (Lacasa, 2008, Lacasa and Luque, 2010, Nu˜nez \textit{et al} 2012). The method that we propose may be summarized as follows.

In the proposed method, we consider one of the time series (e.g. water levels observed in a well) as a reference time series and create time shifted copies of the other time series of interest (e.g. precipitation). We then use VGA to convert all the time series (original, copies and the reference) to graphs and their corresponding adjacency matrices, and compare the copies of the latter time series with that of the reference. The `distance measure' that is used for the comparisons is the usual $L^2$ metric distance (based on the Frobenius norm) between two matrices. We identify the copy of the latter time series for which this distance is minimized compared to the reference, and we define the time shift corresponding to this copy as the time lag between the orginal two time series. More details about VGA and our adaptation to the time lag problem is provided in the next section using mathematical notation. In section 3 we present results from simulations conducted to essentially identify an appropriate sample size and also to assess the performance of the method when values are missing. Section 4 illustrates the application of the proposed method to real hydrogeologic datasets, where we also present a strategy to assess the uncertainty related to the lag estimated. Finally in the last section, we make our concluding remarks.

\section{Method}

Let us denote the two hydrogeological time series that we are interested in, namely precipitation and water levels, by $P(t)$ and $WL(t)$ (or simply $P$ and $WL$), respectively. In order to find the time lag between the two time series, as a first step we fix one of the series, say $WL$, and obtain time-shifted copies of the other series, $P_{\tau_{1}}, \ldots, P_{\tau_{\kappa}}.$ The key step in our methodology is the conversion of all the above time series into graphs based on the visibility graph algorithm. Graphs are mathematical constructs that are used to study relationships among various objects. In graph models the objects of interest are modeled as nodes or vertices and the relationships among the objects are modeled using edges or lines connecting the vertices.\\

\begin{figure}[H]
\begin{center}
\includegraphics[height=3.75in,width=6.5in,angle=0]{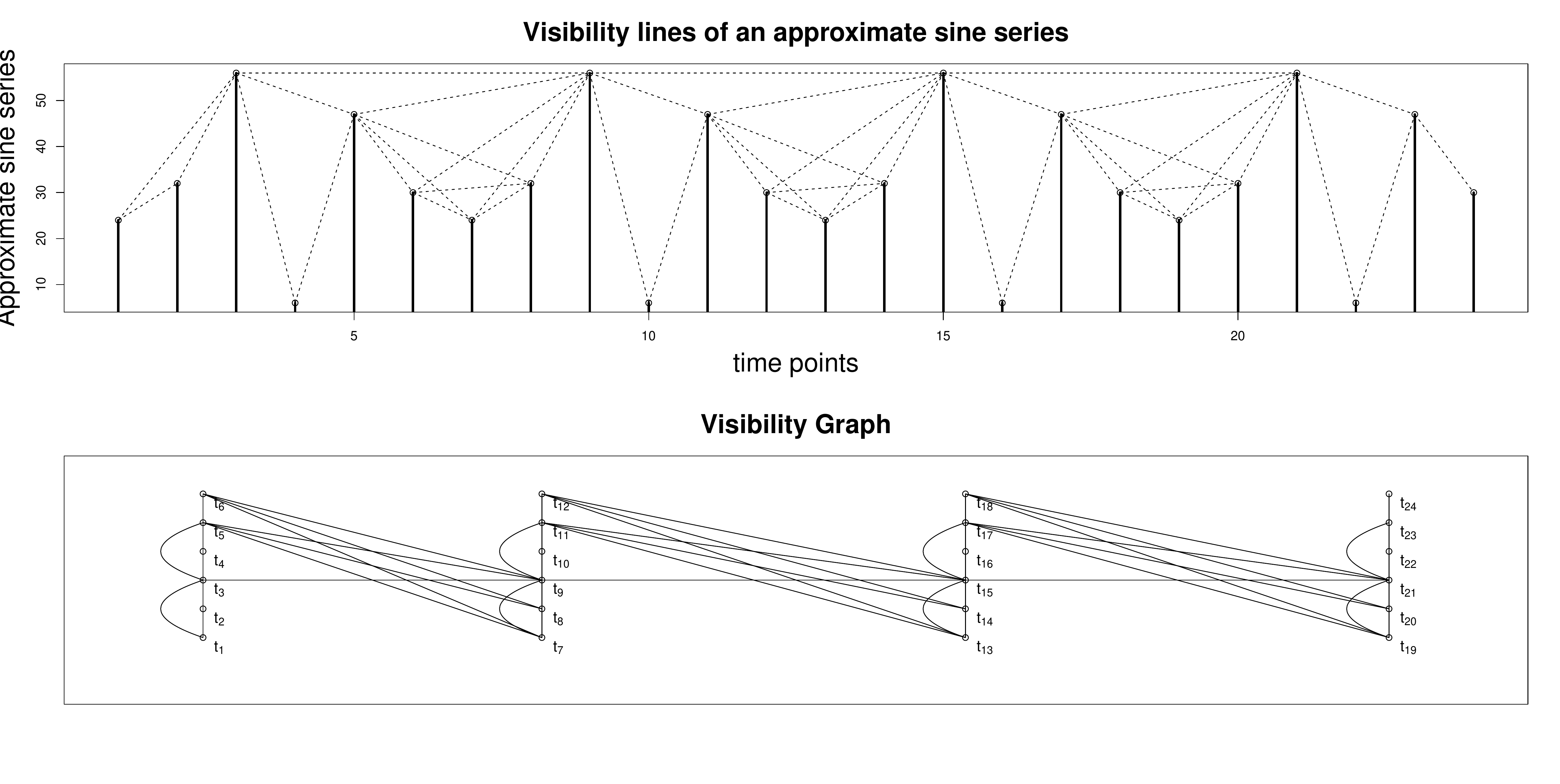}
\caption{A time series and the corresponding visibility graph. $t_{1}, t_{2}$ etc. denote the time points as well as the corresponding nodes in the visibility graph.}
\end{center}
\end{figure}

 Visibility graph algorithm (Lacasa, 2008, Lacasa and Luque, 2010, Nu˜nez \textit{et al} 2012) is a relatively novel method that extends usefulness of the techniques and focus of mathematical graph theory to characterize time series. It has been shown that the visibility graph inherits several properties of the time series, and its study reveals nontrivial information about the time series itself. Figure 1 top panel illustrates how the visibility algorithm works. The time series plotted in the upper panel is an approximate sine series; specifically, a sine series with Gaussian white noise added. The values at 24 time points are plotted as vertical bars. One may imagine these vertical bars as, for example, buildings along a straight line in a city landscape (i.e. a city block). Each node in the associated visibility graph (shown in the bottom panel) corresponds to each time point in the series. So, the graph in Figure 1 has 24 nodes. We draw a link or an edge between a pair of nodes, say $t_{i}$ and $t_{j}$, if the visual line of sight from the top of the building (vertical bar) situated at $t_{i}$ towards the top of the building/bar at $t_{j}$ is not blocked by any intermediate buildings - that is, if we were to draw a line from the top of the vertical bar at $t_{i}$ to the top of the vertical bar at $t_{j}$, it should not intersect any intermediate vertical bars. Visibility lines corresponding to the edges in the graph are plotted as dotted lines in the figure in the upper panel. For example, there is no edge between $t_{2}$ and $t_{4}$ since the line of sight (not shown) between the top points of the vertical bars at these two time points is blocked by the vertical bar at $t_{3}$. On the other hand, there is an edge between $t_{1}$ and $t_{3}$ since the corresponding visibility line (shown as a dotted line) does not intersect the vertical bar at $t_{2}$.\\

 More formally, the following visibility criteria can be established: two arbitrary data values ($t_{q}$, $y_{q}$) and ($t_{s}, y_{s}$) will have visibility, and consequently will become two connected nodes of the associated graph, if any other data ($t_{r}, y_{r}$) placed between them fulfills:
 \[ y_{r} < y_{s} + (y_{q} - y_{s})\frac{t_{s} - t_{r}}{t_{s} - t_{q}}. \]

 This simple intuitive idea has been proven useful practically because of certain nice features exhibited by the graphs generated by this algorithm. First of all they are connected, since each node is connected to at least its neighbors. Secondly, there is no directionality between the edges, so that the graph obtained is undirected. In addition, the visibility graph is invariant under rescaling of the horizontal and vertical axes and under horizontal and vertical translations. In other words, the graph is invariant under affine transformations of the original time series data.\\

 In mathematical notation any graph with $n$ nodes could be represented by its $n \times n$ adjacency matrix $A$ which consists of $0$'s and $1$'s. The $(i,j)^{th}$ element of $A$ is $1$ if there is an edge connecting the $i^{th}$ and the $j^{th}$ node, $0$ otherwise. Two graphs, $G_{1}$ and $G_{2}$, can be be compared by the metric ``distance'', $\|A_{G_{1}} - A_{G_{2}}\|_{2}$ between their corresponding adjacency matrices, $A_{G_{1}}$ and $A_{G_{1}}$. Here, $\|\cdot\|_{2}$, called the Frobenius norm of a matrix, is the square root of the sum of the squares of the elements of the matrix; that is, the square root of the trace of the product of the matrix with itself.\\

 Our proposed method to assess the time lag between the two hydrogeological time series $P$ and $WL$ using the visibility graph approach is as follows: Convert the $WL$ time series into a visibility graph and obtain its corresponding adjacency matrix, $A_{WL}$. Consider time-shifted copies of the $P$ time series, $P_{\tau_{1}}, \ldots, P_{\tau_{\kappa}}$, each shifted in time by a lag from the set $\{\tau_{1},\ldots \tau_{\kappa}\}$. Convert these time-shifted copies of $P$ into their visibility graphs and obtain the corresponding adjacency matrices $A_{P_{\tau_{1}}}, \ldots, A_{P_{\tau_{\kappa}}}$. We determine the copy $A_{P_{\tau_{s}}}$ for which the Frobenius norm $\|A_{WL} - A_{P_{\tau_{s}}}\|_{2}$ is minimized. The time lag between the two original hydrogeological series is then taken as $\tau_{s}$.\\

\begin{figure}[H]
\begin{center}
\includegraphics[height=3.75in,width=6.5in,angle=0]{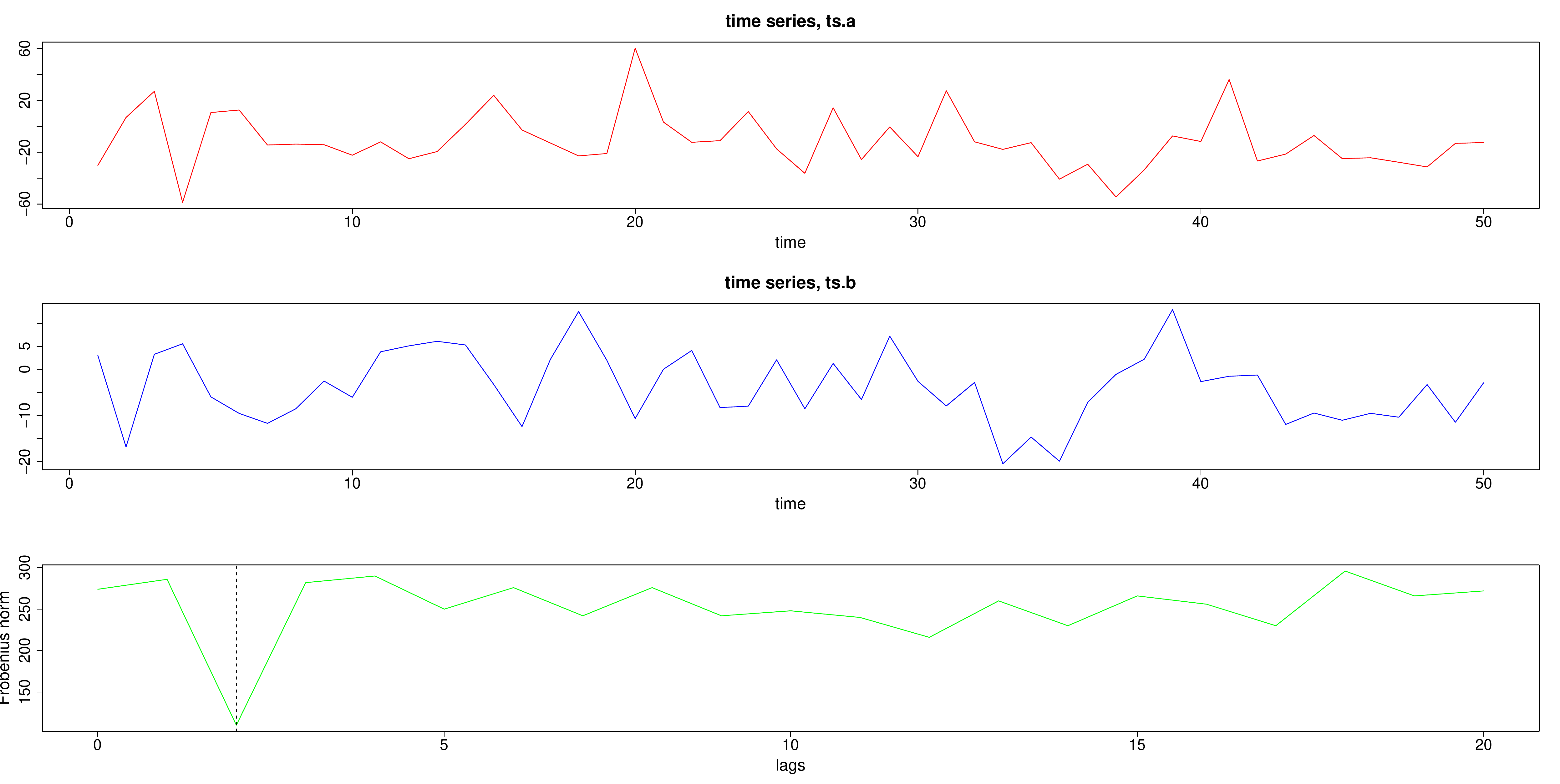}
\caption{Illustration of our method. Top two panels show time series, one shifted by a lag of two from the other. Bottom panel shows the distance-measure based on Frobenius norm for different time lags; minimum is achieved for the time lag 2.}
\end{center}
\end{figure}

We further illustrate our method using the plots in figure 2. The time series in the top panel, $ts.a$ is an approximately a series of values based on the sine function obtained using the following R codes:\\

\noindent \verb"n"  $\leftarrow$ \verb"50" \\
\verb"ts.a" $\leftarrow$ \verb"100*sin(2*pi*(80/1000)*n) + rnorm(n, 0, 25)"\\

The time series, $ts.b$, plotted in the middle panel of Figure 2 is derived from $ts.a$ as follows:\\

\noindent \verb"ts.b" $\leftarrow$ \verb"(1/3)*c(ts.a[3:n], ts.a[1:2]) + rnorm(n, 0, 5)" \\

That is, $ts.b$ is derived by shifting $ts.a$ to the left by two units, by reducing the amplitude to one-third that of $ts.a$, and adding some Gaussian noise. In other words, $ts.a$ and $ts.b$ have roughly the same shape although their amplitudes are different and one is shifted by two time units relative to the other as seen in the figure. One may think of $ts.a$ and $ts.b$ as two time series one affecting the other (since, $ts.b$ is shifted to the left, physically we would think of $ts.b$ affecting $ts.a$); e.g. $ts.b$ as precipitation and $ts.a$ as water levels. Physically, water levels and precipitation never take negative values; so, if one really wants to think of $ts.a$ and $ts.b$ as water levels and precipitation, one could think of them as mean-subtracted and scaled appropriately.

We considered time-shifted copies of $ts.b$ with time-shifts from the following set: $\{0, 1, 2, \ldots, 20\}.$ VGA was applied and adjacency matrices for the corresponding graphs were obtained. Distance-measure based on the Frobenius norm for the time-shifted  copies of $ts.b$ compared to the reference $ts.a$, are plotted in the bottom panel of figure 2. The distance-measure is minimized at 2, which was the lag that we set \textit{a priori}. Thus, in this illustrative example, the lag was correctly identified by the method that we proposed.

\section{Simulations}

\subsection{Sample Size}

We conducted Monte Carlo simulations to assess the performance of the VGA-based method as we varied some of the parameters of the two time series $ts.a$ and $ts.b$ considered in the previous section. The parameters that we considered were \textit{a}) the ratio of the amplitudes between the two simulated series $ts.a$ and $ts.b$, \textit{b}) the variance for the noise term `\verb"rnorm(n, 0, *)"' in the series $ts.a$ (indicated by $*$) and \textit{c}) the variance for the noise term `\verb"rnorm(n, 0, *)"' in the series $ts.b$. For each simulation scenario considered in this section (that is, for each set of the above parameters), 1000 pairs of $ts.a$ and $ts.b$ were generated, and for each pair time lag was assessed based on the proposed method and compared with the lag that was set \textit{a priori}. The performance of the method was assessed based on the percentage of times that the \textit{a priori} lag was correctly identified. The \textit{a priori} lags that we considered for each scenario were $2, 5, 10$ and $15$; we assumed that in typical examples from physical sciences, $2$ will be a small lag and $15$ will be a very large lag.

The reason for considering the ratio of amplitudes was that even if two physical time series (especially, hydrogeological time series) are roughly of the same shape with only a lag between them, their amplitudes (i.e. roughly their 'sizes') are often vastly different. For $ts.a$ and $ts.b$ used in the introductory illustrative example, the ratio of their amplitudes was $1/3$. One of the questions that was addressed in our simulations was whether our method was still good if we changed this ratio drastically, e.g. to $1/9$. Another question that we thought should be addressed is that whether the proposed method works only for nice periodic time series such as the `sine series'. Increasing the variance for the noise term in $ts.a$ makes it less like a 'sine series'. Finally, increasing the variance of the noise term in $ts.b$ makes the shape of $ts.b$ quite different from that of $ts.a$, and by doing so in our simulations we also addressed the performance of the method in such scenarios.

\scriptsize
\begin{table}[!htbp]
\caption{Performance of the method when the ratio of amplitudes between $ts.a$ and $ts.b$ was $1/3$, the noise term in $ts.a$ was $rnorm(n, 0, 25)$ and the noise term for $ts.b$ was $rnorm(n, 0, 5)$.}
\begin{center}
\begin{tabular}{|l|c|c|c|c|}\hline
  Table 1           &  \multicolumn{4}{|c|}{\textit{a priori} set lag}\\\hline
                    &      2      &      5      &    10     &   15    \\\hline
   n = 25           & 94.0\%      &   99.0\%    &  93.0\%   &  97.0\% \\\hline
   n = 50           & 100.0\%     &  100.0\%    & 100.0\%   & 100.0\% \\\hline

\end{tabular}
\end{center}
\end{table}
\normalsize

Our hypothesis was that if we changed the above-mentioned parameters to make the relationship between $ts.a$ and $ts.b$ less ideal than in the illustrative example in the introduction, the performance of the method will be worse. In that sense, essentially the purpose of our simulation was to see whether increasing the sample size will improve the performance in such `bad scenarios', and if so, what would be a recommended minimum sample size that would hedge against such scenarios. In order to do that, we need a reference sample size; that is, a sample size for which the method's performance was excellent when the relationship between $ts.a$ and $ts.b$ was reasonably good (by reasonably good we mean roughly the same shape and size with only a lag in between them). In table 1 above we present the performance of the method for sample sizes $25$ and $50$, when the ratio of the amplitudes and the noise terms were kept exactly the same as in the illustrative example. Since table 1 shows that the performance was excellent for $n = 50$, we consider $50$ as a good sample size choice if we have reasons to believe (may be by visual inspection) that there is a nice relationship between $ts.a$ and $ts.b$.

\scriptsize
\begin{table}[!htbp]
\caption{Performance of the method when sample size is fixed at $n = 50$ and various parameters are changed one at a time from the values for the illustrative example.}
\begin{center}
\begin{tabular}{|l|c|c|c|c|}\hline
   Table 2                       &  \multicolumn{4}{|c|}{\textit{a priori} set lag}\\\hline
                                 &      2      &      5      &    10     &   15    \\\hline
   amplitude ratio = $1/9$       & 71.0\%      &   64.0\%    &  64.0\%   &  64.0\% \\\hline
   $ts.a$ noise: $rnorm(n,0,50)$ & 100.0\%     &  100.0\%    & 100.0\%   & 100.0\% \\\hline
   $ts.b$ noise: $rnorm(n,0,10)$ & 94.0\%      &  92.0\%     &  91.0\%   &  92.0\% \\\hline

\end{tabular}
\end{center}
\end{table}
\normalsize

For the next set of simulations, we fixed the sample size to be $50$, and varied the above-mentioned parameters one at a time. We varied the parameters one a time rather than all together because in the latter case we thought the two time series will be quite unrelated to each other and will not meaningfully represent two physical time series, one affecting the other. The results of this set of simulations are presented in table 2. The first row presents the performance when the noise terms are kept the same as in the introductory illustrative example, but the ratio of the amplitudes was reduced to $1/9$. In this case the performance of the method became drastically worse as seen from the table. In the next row, we present the results when the standard deviation for the noise term for $ts.a$ was changed to $50$ (-for the illustrative example, it was $25$), but the other two parameters were kept the same. This was to check whether the performance became worse if the shape of both time series was not roughly like a sine series. Results from table 2 show that the performance is not affected in this case. These results give reasons to believe that our initial choice of a sine series shape didn't matter; in other words, we would think that the method will perform well no matter what the shapes of the two series are as long as both the series are roughly of the same shape and size. The third row in table 2 shows the results when only the noise term for $ts.b$ was changed. This would correspond to making the shape of $ts.b$ quite different from that of $ts.a$. In this case, the performance is affected  but not very much as the percentages in the third row are all still above $90\%$. Thus, based on table 2, the factors that affected the performance was the ratio of the amplitudes and the noise term for $ts.b$ and among these two, the effect of the former was much more severe than the latter.

\scriptsize
\begin{table}[!htbp]
\caption{Performance of the method when the ratio of amplitudes between $ts.a$ and $ts.b$ was $1/9$, the noise term in $ts.a$ was $rnorm(n, 0, 25)$ and the noise term for $ts.b$ was $rnorm(n, 0, 5)$.}
\begin{center}
\begin{tabular}{|l|c|c|c|c|}\hline
   Table 3           &  \multicolumn{4}{|c|}{\textit{a priori} set lag}\\\hline
                     &      2      &      5      &    10     &   15    \\\hline
   n = 90            &  95.0\%      &  91.0\%    &  91.0\%   &  93.0\% \\\hline
   n = 180           &  98.0\%     &  100.0\%    &  99.0\%   & 100.0\% \\\hline
   n = 365           & 100.0\%     &  100.0\%    & 100.0\%   & 100.0\% \\\hline

\end{tabular}
\end{center}
\end{table}
\normalsize

Next, we wanted to check whether the performance of the method corresponding to the scenario in the first row in table 2 (that is, ratio of amplitudes equals $1/9$ and noise terms for $ts.a$ and $ts.b$ kept exactly the same as for the illustrative example) improved with sample and if so, what could be a recommended minimum sample size. The results for this set of simulations are presented in table 3. As can be seen from the table, performance increases very much when the sample size is increased to $90$, and is near perfect when the sample size is $180$. The percentages for all \textit{a priori} lags are $100\%$ when the sample size is $365$.

\scriptsize
\begin{table}[!htbp]
\caption{Performance of the method when the amplitude for the middle $1/3^{rd}$ section of $ts.a$ was changed from $100$ to $30$. All the other parameters were retained exactly the same as in Table 3, including the ratio of amplitudes between $ts.a$ and $ts.b$ to be equal to $1/9$.}
\begin{center}
\begin{tabular}{|l|c|c|c|c|}\hline
   Table 4           &  \multicolumn{4}{|c|}{\textit{a priori} set lag}\\\hline
                     &      2      &      5      &    10     &   15    \\\hline
   n = 90            &  94.0\%      &  93.0\%    &  94.0\%   &  95.0\% \\\hline
   n = 180           &  99.0\%     &  100.0\%    & 100.0\%   & 100.0\% \\\hline
   n = 360           & 100.0\%     &  100.0\%    & 100.0\%   & 100.0\% \\\hline

\end{tabular}
\end{center}
\end{table}
\normalsize

Finally, we used simulations to check also the performance of the method when the amplitude of the time series varied `seasonally'. It is well-known that the average precipitation varies seasonally; typically, for several consecutive months the average precipitation will be high and for several other consecutive months the average precipitation will be low, and similarly for water levels. In order to mimic this situation somewhat, we generated $ts.a$ series using the same parameters as in table 3, except for the amplitude. We chopped the set of time points into three equal sets of consecutive time points so that, when e.g. $n = 90$, we have the initial $30$ time points $t_{1}-t_{30}$, the middle $30$ time points $t_{31}-t_{60}$ and finally the last $30$ time points $t_{61}-t_{90}$. For the first and last one-thirds of the time points an amplitude of $100$ and for the middle one-third an amplitude of $30$ was used in the simulations for $ts.a$. $ts.b$ was generated with $1/9$ as the ratio of amplitudes, as in table 3. The results for this new set of simulations are presented in table 4. The results look very similar to that seen in table 3, indicating that the additional seasonal variation of the amplitudes did not have any effect on the performance of the method.

The take home message from all the simulations results presented above is that if we are considering hydrogeological time series for which measurements were made daily, then an year's worth of data will be more than sufficient for the proposed method, although the method will work quite well even with $6$ months worth of data. Visually, if the two time series looks clearly to be one affecting the other and of roughly the same shape and size, then even $2$ months worth of data will suffice. Note that in this section, we used the term \textit{sample size} to refer to the number of time points in the time series, and throughout the paper, we implicitly assume that the data points in all the time series are measured at equal time intervals.

\subsection{Missing values}

  Missing values are common in all time series measurements for physical phenomena. In this section we assess, via simulations, the performance of the proposed method in the presence of missing values for two different types of imputation methods - least observation carried forward (LOCF) and mean imputation. In LOCF imputation, if a value is missing at any time point, we carry forward the previous non-missing value; in mean value imputation we impute the average of the prior and the subsequent non-missing values. The missing value mechanism that we considered for our simulations was Missing Completely At Random (MCAR) which means that the missing values are missing exactly as the name implies (completely at random). There is another commonly considered (for example, in clinical studies) missing value mechanism - Missing At Random (MAR) - under which the missingness may depend on the previously observed outcomes. Under this mechanism, both LOCF and mean imputation are known to be biased. But, we consider that the missingness in hydrogeological time series do not depend on the previously observed outcomes, and hence the MAR assumption is unrealistic, and thus we did not consider MAR for our simulations. As a matter of fact, the missingness that we have seen for real hydrogeological time series is as follows - for example, for a time series in which measurements are made daily, non-missing measurements are seen for a large chunk of consecutive time points (6 months to 2 years) followed by a large chunk of data missing at a stretch (several weeks or months), which again is followed by a large chunk of non-missing data, and so on. When a chunk of data is missing for a large number of consecutive time points, none of the existing imputation methods will work very well. In such cases, the best strategy is to analyze separately the large chunks of data with no missing values at all. Nevertheless, we conducted the following simulations for hypothetical scenarios.

  In all the simulations reported in this section, we fixed the sample size to be 180, and we used the same noise terms for $ts.a$ and $ts.b$ as in the illustrative example in the introduction. The ratio of amplitudes was set to be $1/9$ as in the simulations for table 3. In order to adhere to the MCAR mechanism we randomly set either 9 or 18 or 27 or 36  values to be missing; 9, 18, 27 and 36 correspond to $5\%, 10\%, 15\%$ and $20\%$ of 180. Furthermore, we considered scenarios where the values were set to be missing for only one time series ($ts.a$) or for both. If it were set to be missing for both, then it was at the same time points for both, which we think is the more realistic scenario.

\scriptsize
\begin{table}[!htbp]
\caption{Performance of the method with imputation methods for missing values. The sample size was fixed to be 180. All the other parameters were retained exactly the same as in Table 3, including the ratio of amplitudes between $ts.a$ and $ts.b$ to be equal to $1/9$.}
\begin{center}
\begin{tabular}{|l|c|c|c|c|c|}\hline
   Table 5                    &   \multicolumn{5}{|c|}{}                                                            \\\hline
   \textit{a priori} set lag  &  No. missing   & \multicolumn{2}{|c|}{LOCF}    & \multicolumn{2}{|c|}{Mean Imputation}  \\\cline{3-6}
                              &                &   both TS      &  only 1 TS   &   both TS   &  only 1 TS               \\\hline
   \multirow{4}{*}{2}         &   9            &  97.0\%        &   99.0\%     &  100.0\%    & 100.0\%                  \\\cline{2-6}
                              &   18           &  92.0\%        &   97.0\%     &   95.0\%    &  98.0\%                  \\\cline{2-6}
                              &   27           &  91.0\%        &   96.0\%     &   93.0\%    &  98.0\%                  \\\cline{2-6}
                              &   36           &  84.0\%        &   92.0\%     &   82.0\%    &  92.0\%                  \\\hline\hline
   \multirow{4}{*}{5}         &   9            & 100.0\%        &   99.0\%     &   99.0\%    & 100.0\%                  \\\cline{2-6}
                              &   18           &  94.0\%        &   98.0\%     &   94.0\%    &  98.0\%                  \\\cline{2-6}
                              &   27           &  95.0\%        &   95.0\%     &   91.0\%    &  97.0\%                  \\\cline{2-6}
                              &   36           &  86.0\%        &   94.0\%     &   90.0\%    &  98.0\%                  \\\hline\hline
   \multirow{4}{*}{10}        &   9            & 100.0\%        &   99.0\%     &   98.0\%    &  99.0\%                  \\\cline{2-6}
                              &   18           &  97.0\%        &   98.0\%     &   97.0\%    &  99.0\%                  \\\cline{2-6}
                              &   27           &  92.0\%        &   96.0\%     &   92.0\%    &  97.0\%                  \\\cline{2-6}
                              &   36           &  77.0\%        &   93.0\%     &   81.0\%    &  94.0\%                  \\\hline\hline
   \multirow{4}{*}{15}        &   9            &  98.0\%        &  100.0\%     &  100.0\%    & 100.0\%                  \\\cline{2-6}
                              &   18           &  99.0\%        &   98.0\%     &   98.0\%    &  97.0\%                  \\\cline{2-6}
                              &   27           &  84.0\%        &   96.0\%     &   92.0\%    &  98.0\%                  \\\cline{2-6}
                              &   36           &  79.0\%        &   94.0\%     &   88.0\%    &  94.0\%                  \\\hline
\end{tabular}
\end{center}
\end{table}
\normalsize

  The performance of the proposed method with both LOCF and mean imputation was near perfect when only $5\%$ of the values (that is, 9 out of 180) were missing. This was true regardless of whether the values were missing for only one time series or for both, and also true across all \textit{a priori} set lags, 2, 5, 10 and 15. When $10\%$ of the values (that is, 18 out of 180) were missing for only one time series, the method did very well under both LOCF and mean imputation for all lags. When $10\%$ of the values were missing for both time series, the performance was still very good when the lags were large (10 or 15); when the lags were small (2 or 5), the performance with both imputation methods was still good but not as good as when the lags were large. For example, when $10\%$ values were missing and when the lag was 2, the performance with LOCF was $97\%$ and $92\%$, respectively, depending on whether the values were missing for only one time series or both; the corresponding values for lag 10, on the other hand, were even better: $98\%$ and $97\%$.

  With $15\%$ missing values (27 out of 180), the performance was still good (that is, in the range $90\%-97\%$) with LOCF and mean imputation, for lags 2, 5, and 10, irrespective of whether it was missing for only one or for both time series (although, of course, if it was missing only for one time series, it was better). However, when the \textit{a priori} set lag was 15, the performance with LOCF was weak ($84\%$), when $15\%$ values were missing for both time series; it was still good ($96\%$) with LOCF when only one time series had $15\%$ missing values, and with mean imputation also ($92\%$ and $98\%$). With $20\%$ missing values the method worked well under both types of imputations and for all lags, only when one time series had missing values. When both time series had $20\%$ missing values, the performance of LOCF was not good with small lags ($84\%$ for lag 2 and $86\%$ for lag 5) and got worse for larger lags ($77\%$ for lag 10 and $79\%$ for lag 15). The performance with mean imputation was slightly better ($82\%, 90\%, 81\%$ and $88\%$, for lags 2, 5, 10 and 15, respectively) but still not quite up to the mark.

  In summary, based on the above simulation results, we consider it quite safe to use the proposed method in conjunction with either of the imputation methods if it is only $5\%$ values missing for only one time series or for both. With $6$-$15\%$ values missing, the imputation methods give good results only if it is missing for one time series; if it is missing for both, then it is not very safe to say that imputations will work, but still reasonably safe. With about $20\%$ of the values missing for both time series, it is definitely not recommended to use the proposed method with either of the imputations although it may be somewhat acceptable if it is missing for only one time series. Also, in general, we observed that the performance with mean imputation was slightly better except for one or two scenarios. If the statistical practitioner has a preference of one method over the other, it may still be recommended to use both for the proposed method, at least as a sensitivity analysis. Finally, we emphasize again the point made in the beginning of the section, that if large chunks of data are missing at a stretch then the imputation methods won't work; in such cases, it is better to focus the analysis on other chunks of data with no or very sparse missing values.

\section{Real data example}

  In this section, we present the results from an application of the proposed method on real hydrogeological time series. In Southwest Florida, two of the shallow aquifers that are tapped for water supply are the unconfined Water Table Aquifer and the semi-confined Lower Tamiami Aquifer. These aquifers are considered as sources of limited supply and regulated by the South Florida Water Management District (SFWMD, 2015a). Water Table Aquifer is generally less than 50 feet thick and consists of less permeable unconsolidated sediments at the upper portion and relatively permeable limestone at the basal portion. The Water Table Aquifer is hydraulically separated from the Lower Tamiami Aquifer by about 15 to 30 feet of confining beds that consists of low permeable strata (Bonita Springs Marl Member and Caloosahatchee Clay Member). The top of Lower Tamiami Aquifer is generally between 60 and 80 feet below land surface.  This aquifer extends to 100 to 150 feet below grade and generally consists of sandy, biogenic limestone and calcareous sandstone (SFWMD, 2015b). This aquifer is one of the primary sources of public water supply in southwest Florida. To understand the lag responses of rainfall in these shallow aquifers are important for water management. For this study, in order to determine the lag responses within these two aquifers due to rainfall events, we utilized daily water level data recorded in the Water Table Aquifer and Lower Tamiami Aquifer.

  It is relevant to note that data collected in shorter frequencies (e.g. hourly) are ideal for “lag” related studies. Hourly water level data was available from the Water Table Aquifer well as well as the Lower Tamiami Aquifer in the study area; however, precipitation data was available only on a daily basis. In order to have both water level data and precipitation in the same time interval, we averaged the water levels to a daily average. The daily-averaged data was used solely for illustration of the statistical technique presented in this paper. Data was available from July 1st, 2010 till June 29th 2016, but water level data were missing for the following time intervals: January 5th, 2011 to April 8th, 2012, July 10th, 2012 to October 1st, 2012, April 2nd, 2013 to September 30th, 2013, and finally between April 2nd, 2014 to June 29th, 2014. Complete data was available between June 30th, 2014 and June 29th, 2016 (731 days); we analyzed this data for our illustration since this was the largest available time period with no missing data.  The water level and precipitation data that were analyzed are graphically presented on Figure 3.

 \begin{figure}[H]
    \begin{center}
    \includegraphics[height=3.75in,width=6.5in,angle=0]{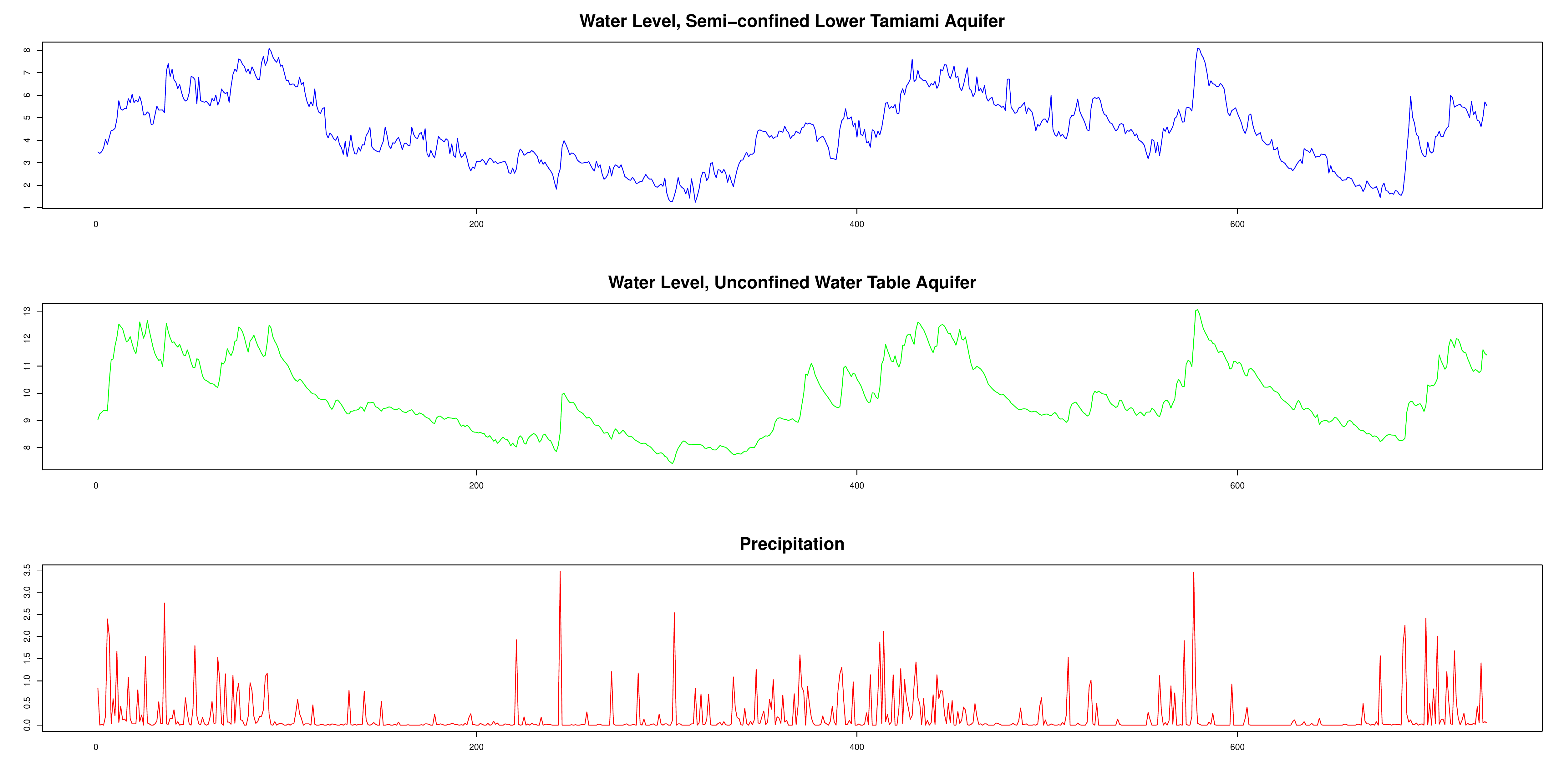}
    \caption{Time series plots of the real data: daily water levels for the aquifers (top two panels) and daily precipitation (bottom panel) }
    \end{center}
  \end{figure}

  VGA method was applied to all the times series plotted in figure 3, and Frobenius distance between the corresponding pairs of adjacency matrices are plotted in figure 4. For both unconfined Water Table Aquifer and semi-confined Lower Tamiami Aquifer, the Frobenius distance is minimized at lag 2. This makes hydrogeological sense since, although one aquifer is a bit deeper than the other, considering the difference in total depths between the wells is roughly only about 40 feet, the water level response in the relatively deeper well may be observed in a matter of hours. In figure 4, for both plots, we note that although the Frobenius distnace for lag 1 is not the minimum it is close to the minimum compared to that of the other lags. Thus, in order to check whether the minimum was attained at lag 2 just due to chance, we need to quantify the uncertainty regarding the estimate. Since naive resampling strategies like bootstrap would create independent samples (that is, with autocorrelations near zero), we used the following subsampling strategy that would preserve the original autocorrelations. We set a time-window-size, say 100, and use the consecutive data points for water levels and precipitation in that window, and apply the proposed method to this sub-sampled data, as we did for the original data. First we place the left end of the window on the first time point of the original data, conduct the analysis, find the lag, and then move the window one unit to the right and repeat the analysis. We continue this process iteratively until the right end of the window touches the final time point for the original data. Thus, with a window-size of 100, and with a time series of length 731, we will have 631 iterations, and for each iteration, we will have a lag between the pair of time series under consideration, so that we will have 631 lags at the end.

 \begin{figure}[H]
    \begin{center}
    \includegraphics[height=3.75in,width=6.5in,angle=0]{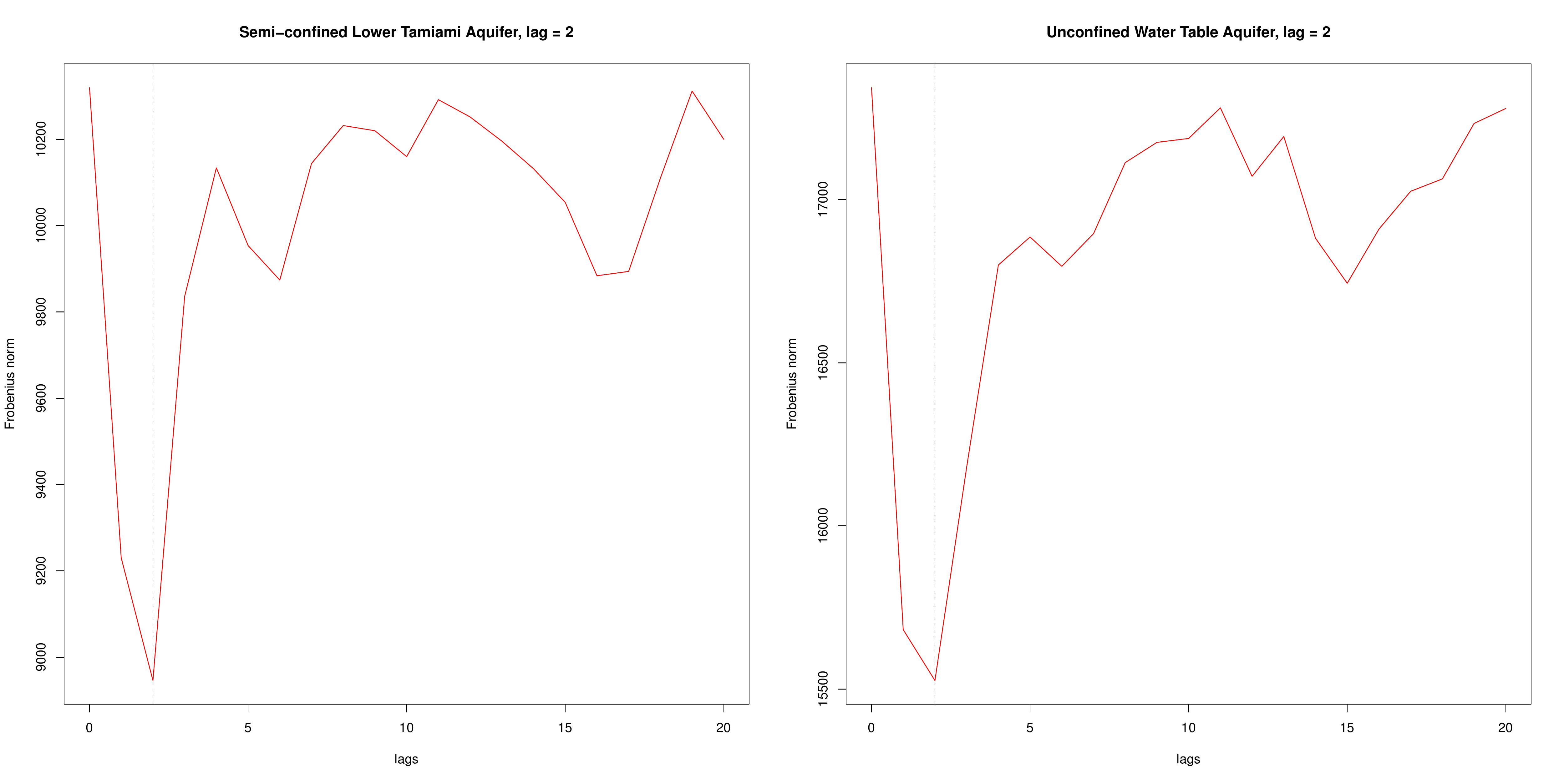}
    \caption{Distance-measure based on Frobenius norm for different time lags for semi-confined Lower Tamiami Aquifer (top panel) and unconfined Water Table Aquifer (bottom panel)}
    \end{center}
  \end{figure}

  The histograms for the 631 lags obtained using this iterative process are plotted in the top panel of figure 5. The highest frequency for the Lower Tamiami Aquifer is at lag 2 consistent with our finding for the original time series with 731 points. However, interestingly, the highest frequency for the Water Table Aquifer was at lag 1, although the frequency for lag 2 is almost as high. Now the question arises whether this reversal was due to the size of the window (100), which is quite smaller than the length of the original series (731). So, we repeated the analysis with a window-size of 365 (that is, 366 iterations). The results of the second iterative analysis are shown in the bottom panel of figure 5; in this case, the highest frequency for both aquifers is at lag 2. We repeated the analysis using a window size of 50 (681 iterations) and 25 (706 iterations); in these analyses, only lag 2 appeared for all windows, so that the histogram will look like a single bar at 2 (of height 50 or 25, respectively), and no bars at any other lags. Since this is simple enough to convey without a histogram, we didn't plot the histograms for window-sizes 50 and 25. Based on these results, and on hydrogeological sense, we would conclude that on the average, the water levels rise and fall 2 days after a corresponding fluctuation in precipitation, for both the aquifers. The analysis presented in this section suggests that it is critical to quantify the uncertainty prior to making conclusions and that the selected window size can influence the conclusions.

 \begin{figure}[H]
    \begin{center}
    \includegraphics[height=3.75in,width=6.5in,angle=0]{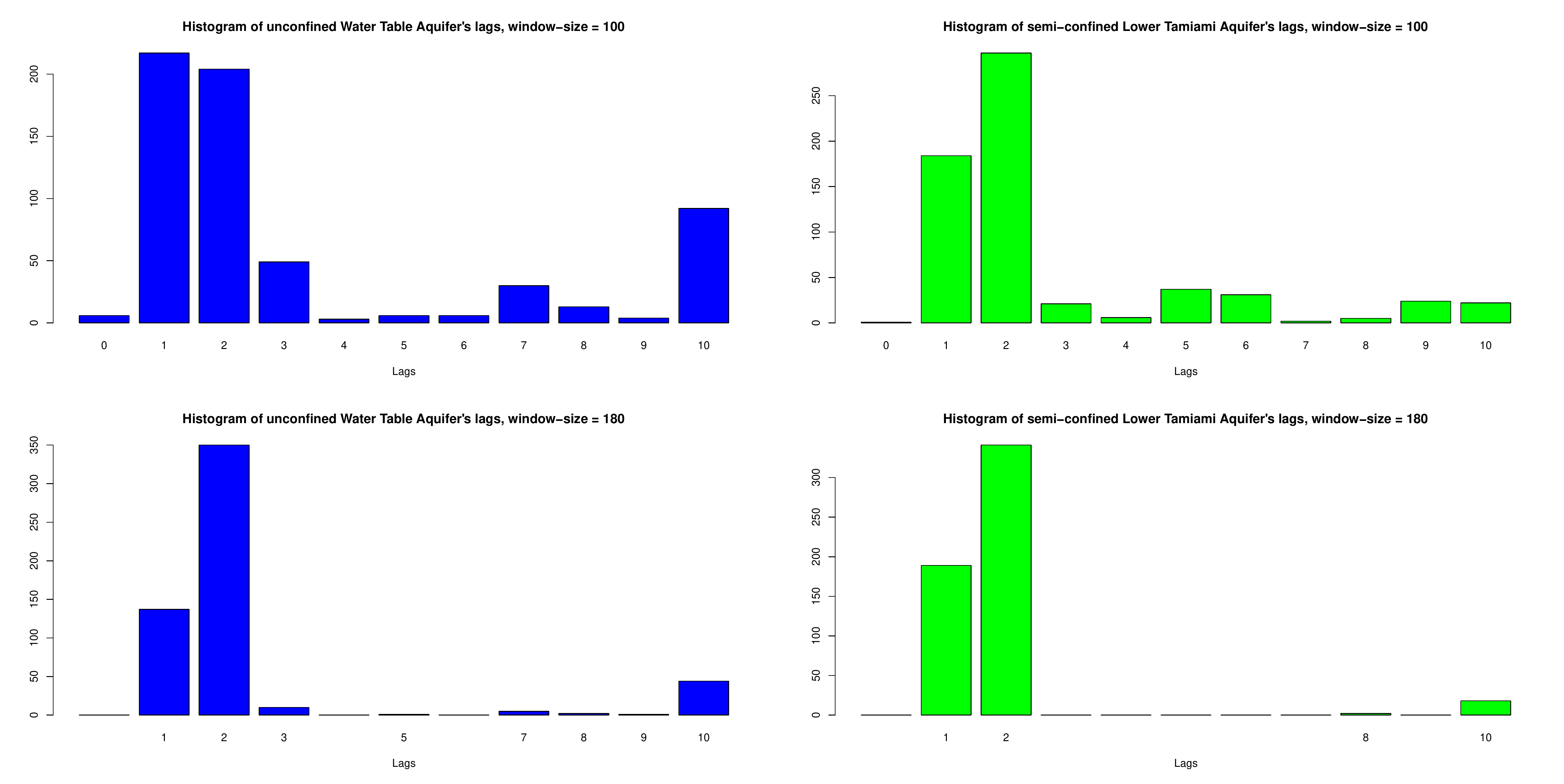}
    \caption{Histograms of the lags obtained for smaller datasets obtained using a window size of 100 (top panel) and a window size of 180 (bottom panel). The figures on the left correspond to lags for unconfined Water Table Aquifer, and on the left correspond to semi-confined Lower Tamiami Aquifer.}
    \end{center}
  \end{figure}

\section{Conclusions} Quantifying time lags between two time series data, where one affects the other, is important for modelers of many physical phenomena, especially in hydrogeology. We propose an approach based on a simple extension of the visibility graph algorithm. We conducted simulations to assess the performance of the proposed method under different scenarios, and determined that the method worked well under reasonable settings. Based on simulations we were also able to recommend sample size necessary to conduct the proposed analysis, and the maximum percentage of missing values under which the method will still work reasonably well with imputations. We also illustrated the method by applying it real data of water levels from aquifers and precipitation levels, and emphasized the importance of quantifying the uncertainty related to estimate of the lag.


\end{document}